\documentclass{jltp}
\usepackage{graphicx} % uncomment this line to include the graphicx package
\newcommand \beq{\begin{eqnarray}}
\newcommand \eeq{\end{eqnarray}}
\newcommand \bea{\begin{eqnarray}}
\newcommand \eea{\end{eqnarray}}

\newcommand \la{\raisebox{-.5ex}{$\stackrel{<}{\sim}$}}

\title{Rapidly Rotating Bose-Einstein Condensates}

\author{Gordon Baym}

\address{Department of Physics, University of Illinois at
Urbana-Champaign\\ 1110 West Green Street, Urbana, Illinois 61801}

\runninghead{G. Baym}{Rapidly Rotating Bose-Einstein Condensates}

\begin{document}

\maketitle

\begin{abstract}

    How does a rapidly rotating condensed Bose gas carry extreme amounts of
angular momentum?  The energetically favored state of a not-too-rapidly
rotating Bose condensed gas is, as observed, a triangular lattice of singly
quantized vortices.  This paper describes the fates of the vortex lattice in
both harmonic and anharmonic traps when condensates are rotated extremely
rapidly.


\end{abstract}

\section{INTRODUCTION}

    Superfluids respond to rapid rotation by forming triangular arrays of
singly quantized vortex lines.  Compared with superfluid $^4$He,\cite{packard}
the properties of vortices in Bose-Einstein condensed atomic gases have proven
to be considerably more accessible.  Following the creation of single vortex
lines in atomic condensates by phase imprinting,\cite{JILA} large arrays have
been created via mechanical rotation, by stirring the condensate with a laser
beam,\cite{Madison,Abo} and by particle evaporation.\cite{HaljanCornell}
Upwards of 300 vortex lines have been created in rapidly rotating condensates,
and properties such as the modes of the lattice\cite{jilatk,jila3} and vortex
core structure\cite{jila3,coddington} have been studied in detail.

    The superfluid velocity, $\vec v$\,, obeys the quantization condition,
$\oint \vec v\cdot d\vec\ell = (h/m)N_v(\cal{C})$, where the line integral is
along a closed contour surrounding $N_v(\cal{C})$ singly quantized vortices,
$h$ is Planck's constant, and $m$ the particle mass.  In the neighborhood of a
single line, the velocity is in the azimuthal direction and has magnitude
$\hbar/m\rho$, where $\rho$ is the distance from the line.  A system
containing many vortex lines appears to rotate uniformly with an average
angular velocity $\Omega$, which is simply related to the (two-dimensional)
density of vortex lines, $n_v$, by the quantization condition:  $\Omega =
\pi\hbar n_v/m$.  As the system rotates more and more rapidly, the lines
become closer and closer.  The question I would like to discuss here is what
eventually happens to the vortex array.  In other words, how does a rapidly
rotating Bose condensate carry large quantities of angular momentum per
particle?

    One may be tempted to argue by analogy with a Type II superconductor which
in the presence of a magnetic field above a critical value, $H_{c1}$, contains
an array of vortex lines with quantized flux.  As the field is increased the
line density grows until the cores begin to overlap, at a critical field,
$H_{c2}$, at which point the system turns normal.  However, a low temperature
rotating bosonic system {\em does not} have a normal phase to which it can
return.\cite{Fetter} Unlike in a weakly interacting superconductor, where
condensation occurs as a small dynamical decoration on top of a normal state,
Bose condensation occurs kinematically.  A Bose system must respond
differently than a Type II superconductor.  In He II, the question has never
been of experimental interest, since the intervortex spacings are typically
macroscopic, while the core sizes are of order {\AA}ngstroms, and thus the
critical rotation rate, $\Omega_{c2}$, at which the vortex cores would
approach each other is unobservably large, $\sim 10^{12}$
rad/sec.\cite{duncan} In a weakly interacting atomic condensate, the radius of
a single vortex core is of order $\xi_0 = 1/\sqrt{8\pi n a}$, where $a$ is the
s-wave scattering length, and $n$ is the particle density.  Typically, $\xi_0
\sim 0.2$ $\mu$m, so that cores would begin to touch at $\Omega_{c1}\sim
8\pi^2 n a\hbar/m \sim 10^3 - 10^5$ rad/sec, an experimentally
accessible rate.

    The ground state of the system rotating at angular frequency $\Omega$
about the $z$ axis is determined by minimizing the energy in the rotating
frame, $E'=E - \Omega L_z$, where $L_z$ is the component of the angular
momentum of the system along the rotation axis:
\begin{eqnarray}
 E'& = &
  \int d^3 r \left[\frac{\hbar^2}{2m} \left|\left(
   -i\nabla  -m {\vec\Omega} \times {\vec r}\right)\psi\right|^2 \right.
     \nonumber\\
   & & \qquad\quad\left.
   +\left(V(\vec r\,)-\frac12 m \Omega^2 r_\perp^2\right) |\psi|^2 + \frac12 g
|\psi|^4\right]
\label{Eprime}
\end{eqnarray}
where $\psi$ is the condensate wave function, $\vec r_\perp = (x,y)$, $V(\vec
r\,)$ is the trapping potential, and $g=4\pi\hbar^2a/m$ determines the
strength of the interparticle interaction.

    The fate of a rapidly rotating Bose system depends on how the system is
confined.  In typical condensate experiments the system, rotating about the
$z$ axis, is confined in a harmonic trap of the form,
\begin{equation}
  V(r_\perp,z)
= \frac12 m\left(\omega_\perp^2r_\perp^2 + \omega_z z^2\right).
\label{harmonic}
\end{equation}
In this case the centrifugal potential, $-\frac12 m \Omega^2 r_\perp^2$,
tends to cancel the transverse trapping potential, and the system cannot
rotate faster than $\omega_\perp$ without becoming untrapped.  As $\Omega \to
\omega_\perp$, the system flattens out, becomes almost two dimensional, and
eventually enters quantum Hall-like states.  On the other hand, in the
presence of an anharmonic transverse trapping potential that grows faster than
quadratic, e.g.,
\begin{equation}
  V_\perp(r_\perp)
  = \frac12 m\omega_\perp^2r_\perp^2(1+\lambda r_\perp^2),
 \label{anharmonic}
\end{equation}
the system can rotate faster than $\omega_\perp$, since it is contained by
the anharmonic part of the potential.

\section{THE VORTEX CORES SHRINK WITH INCREASING $\Omega$}

    In fact there is never a phase transition associated with the vortex cores
overlapping in a rotating Bose condensate.  Rather, as the intervortex spacing
becomes comparable to the mean field coherence length, $\xi_0$, the vortex
cores begin to shrink, and eventually the core radius scales down with the
intervortex spacing.

\begin{figure}[htbp]
\centerline{\includegraphics[height=2.5in]{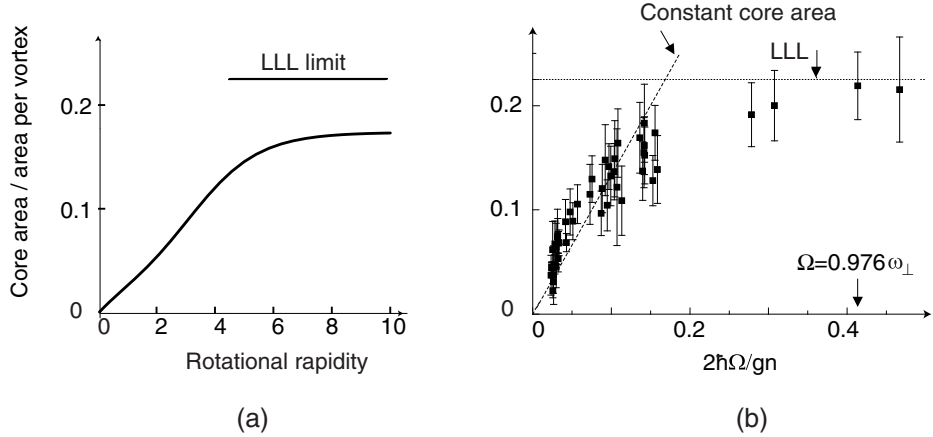}}
%
%\framebox[5in]{\rule[1.125in]{0in}{1.125in}}
%\makebox[5in]{\rule[1.125in]{0in}{1.125in}}
\caption{(a) Mean vortex core area as a fraction of the area per vortex
vs. the rotational rapidity (see text).  (b) Measured vortex core area as a
fraction of the area per vortex vs. 2$\Omega/gn$.  Adapted from
Ref.~\onlinecite{jila3}.}
\label{fig:core}
\end{figure}

    In Ref.~\onlinecite{FG}, and more fully in Ref.~\onlinecite{coresize}, we
approached the problem by treating the core radius, $\xi$, as a variational
parameter.  By integrating out the short range structure on scales of the
vortex separation, one derives an effective Hamiltonian, \begin{eqnarray} E'
&=& \int d^3r\, n(\vec r) \left[-\frac{\Omega^2}{2}mr^2 + V(\vec r\,) +
a\hbar\Omega + \frac12 bgn(\vec r\,)\right], \label{Erot} \end{eqnarray} where
$n(\vec r)$ is the smoothed density; the parameter $a(\xi)$ takes into account
the local vortex kinetic energy, including the curvature of the order
parameter, and $b(\xi)$ describes the renormalization the coupling constant,
$g$, due to density fluctuations about the smoothed density within each vortex
cell.  The core size is determined by minimizing Eq.~(\ref{Erot}) with respect
to $\xi$.  Figure 1a shows the resultant mean square area of the vortex core,
in a harmonic trap, measured in units of the area per vortex line, for the
qualitatively accurate model in which the condensate wave function has the
form $\psi(\rho) \sim \rho$, for $\rho<\xi$, and constant for $\xi<\rho<\ell$,
where $\ell$ is the radius of the (cylindrical) Wigner-Seitz cell around a
given vortex:  $\ell^2 = 1/m\Omega$.  In this model, the mean square core area
divided by the area per vortex, $\cal{A}$, is $\xi^2/3\ell^2$.  The horizontal
axis in Fig. 1 is the rotational rapidity,\cite{coresize} $y$, defined by
$\Omega/\omega_\perp = \tanh y$, a convenient variable for spreading out the
region where $\Omega\, \la\, \omega_\perp$.  The linear rise at small $\Omega$
occurs because the core size remains constant, while $\ell^2$ decreases
linearly with $1/\Omega$.  The flattening of $\cal{A}$ with increasing
$\Omega$ is a consequence of the vortex radius scaling with the intervortex
spacing.  The upper line shows the exact limit as $\Omega\to\omega_\perp$.
Recent JILA measurements\cite{jila3,coddington} of $\cal{A}$ (as a function of
$2\hbar\Omega/gn$), Fig. 1b, nicely show the expected initial linear rise,
followed by the predicted scaling of the core radius with intervortex spacing.

\section{LOWEST LANDAU LEVEL REGIME}

    As the rotation rate in a harmonic trap approaches $\omega_\perp$, the
centrifugal potential basically cancels the transverse trapping potential; the
cloud flattens out, and becomes a effectively two dimensional system.  Because
the density $n$ of the system drops, the interaction terms $\sim gn$, become
small compared with $\hbar\Omega$, Then, as one sees from Eq.~(\ref{Eprime})
with (\ref{harmonic}), the dynamics is that of a particle feeling the Coriolis
force alone, a system formally analogous to a particle in a magnetic field.
Ho\cite{Ho} predicted that in this limit particles should condense into the
lowest Landau level (LLL) in the Coriolis force, similar to charged particles
in the quantum Hall regime.  When $2gn \ll \Omega$, the states in the next
higher Landau level are separated by a gap $\simeq 2\omega_\perp$.  This
insight has led to extensive experimental studies in which rotation rates in
excess of 0.99 $\omega_{\perp}$ have been achieved.\cite{HaljanCornell,jila3}

    The single particle wave functions in the lowest Landau level have the
form, $\phi_{\mu}({\vec r_\perp}) \sim \zeta^\mu e^{-r_\perp^2/2d_{\perp}^2}$,
where $\zeta = x+iy$, $\mu = 0,1,2,\dots$, and the transverse oscillator
length, $d_\perp$, is given by $\sqrt{\hbar/m\omega_\perp}$.  The LLL
condensate wave function is a linear superposition of such states:
\begin{equation}
  \phi_{\rm LLL}({\vec r_\perp}) \sim \Sigma_\mu c_\mu \zeta^\mu
    e^{-r^2/2d_{\perp}^2} \sim \prod_{i=1}^{N}(\zeta-\zeta_i)\
   e^{-r^2/2d_{\perp}^2},
 \label{LLL}
\end{equation}
where the polynomial $\Sigma_\mu c_\mu \zeta^\mu$ is written as a product
over its zeroes, $\zeta_i$, which are simply the positions of the vortices in
the condensate.  The state (\ref{LLL}) in the regime $gn \ll \hbar\Omega$ is
a direct continuation of the state in the slowly rotating regime,
$\hbar\Omega \ll gn$.

\section{DENSITY PROFILE AND LATTICE DISTORTION}

    As long as the total number of vortices is much larger than unity, the
energy of the cloud in the LLL limit is given by\cite{gentaro}
\begin{equation}
   E'= \Omega N
   +\int d^3r\{(\omega_{\perp}-\Omega) \frac{r_\perp^2}{d_{\perp}^2}n(\vec
    r\,) +\frac{bg}{2} n(\vec r\,)^2\},
\label{k2}
\end{equation}
plus terms involving the trapping potential in the $z$ direction.  Here
$n(\vec r\,)$ is the smoothed density profile, $=\langle |\phi_{\rm
LLL}|^2\rangle$; the brackets denote the long wavelength smoothing.  The
energy (\ref{k2}) is minimized when the cloud assumes a density profile of the
Thomas-Fermi form,\cite{BP} $n(r_\perp)\sim (1-r_\perp^2/R^2)$ -- an inverted
parabola -- where $R$ is the transverse radius of the cloud.  For $Na/d_z
\gg 1$, where $d_z$ is the axial oscillator length, the structure in the
radial direction will be Thomas-Fermi at large $\Omega$, even if it is
Gaussian at small $\Omega$.\cite{coresize} In
experiment,\cite{jila3,coddington} the density profile indeed remains an
inverted parabola as $\Omega \to \omega_\perp$.

    Since the energy (\ref{k2}) depends only on the smoothed density, the
vortices must adjust their locations in order that the smoothed density be an
inverted parabola.  From the arguments in Ref.~\onlinecite{Ho}, we find the
relation between the smoothed density and the mean vortex density,
$n_v(r_\perp)$,
\begin{equation}
 \frac{1}{4}\nabla^2 \ln n(r_\perp)  = -\frac{1}{ d_\perp^2} +\pi n_(r_\perp).
  \label{nv}
\end{equation}
For a Gaussian density profile, the vortex density is constant.  However, for
a Thomas-Fermi profile,
\begin{equation}
   n_v(r_\perp) = \frac{1}{\pi d_\perp^2} -
  \frac{1}{\pi R^2}\frac{1}{\left(1-r_\perp^2/R^2\right)^2}.
  \label{nnv}
\end{equation}
Since the second term is of order $1/N_v$ compared with the first, the
density of the vortex lattice is basically uniform (and the vortex array forms
an almost perfect triangular lattice).  Turning the argument around, very
small distortions of the vortex lattice from triangular can result in large
changes in the density distribution.\cite{distort} Recent measurements of the
(percent scale) distortions of the vortex lattice at relatively low
rotation rates,\cite{coddington} are in good agreement with theory.

\section{LATTICE MODES}

    The collective modes of the vortex lattice in a rotating superfluid have
been of considerable interest since the 1960's, when Tkachenko\cite{Tkachenko}
showed that the lattice supports an elliptically polarized mode, a mode
observed in helium in 1982.\cite{andereck} To describe the Tkachenko modes,
Ref.~\onlinecite{BC} reformulated the hydrodynamics of rotating superfluids to
take into account the elasticity of the vortex lattice (including normal
fluid, dissipation, and line bending -- Kelvin -- oscillations of the vortex
lines in three dimensions); effects of the oscillations of the vortex lines at
finite temperature on the long range phase correlations of the superfluid were
discussed in Ref.~\onlinecite{lattice}.  In superfluid helium, where $\Omega$
is always much smaller than characteristic phonon frequencies, $sk \sim s/R$,
where $s$ is the sound velocity and $R$ the transverse size, the modes
frequencies are linear in the wave vector, $\omega_T = (2C_2/mn)^{1/2}k$,
where $C_2$, the shear modulus, equals $\hbar n\Omega/8$.  However, in Bose
condensates it is possible to go to the rapidly rotating
regime\cite{jilatk,jila3,coddington} where $\Omega > sk$.  For general
$\Omega$, the in-plane Tkachenko mode frequencies are given
by,\cite{tkmodes,qhmodes}
\begin{equation}
  \omega_{T}^2 = \frac{2C_2}{nm} \frac{s^2k^4}{\left\{4\Omega^2 +
   (s^2+4(C_1+C_2)/nm)k^2\right\}},
  \label{tk}
\end{equation}
where $C_1$ is the compressional modulus of the vortex lattice.  Note that
at low $k$, the mode frequency becomes $\sim k^2$.  In addition the system
supports a gapped sound mode of frequency,
\begin{equation}
  \omega_I^2 = 4\Omega^2 + \left(s^2+\frac{4}{nm}(C_1+C_2)\right)k^2.
 \label{inertial}
\end{equation}
Figure~2a shows the frequencies as a function of k, while Fig.~2b shows
the lowest mode frequency (of wave vector\cite{crescimanno} $5.45/R$)
evaluated for a trapped rotating $^{87}$Rb gas of $N=2.5\times 10^6$
particles.\cite{cps} In experiment,\cite{jilatk,jila3} the particle number
varies from run to run; here we have multiplied the experimental frequencies
by a factor $(N/2.5\times 10^6)^{-1/5}$ to compare with theory, calculated for
$N=2.5\times 10^6$.  As one approaches the LLL regime, the shear modulus
decreases to\cite{qhmodes,SHM} $C _2 \simeq (81/80\pi^4)ms^2 n$; the
Tkachenko
modes are then softer by a factor $(9/\pi^2)(ms^2/10\Omega)^{1/2}$; see solid
curve in Fig. 2a.  This softening of the modes has recently been
measured,\cite{jila3,coddington} and is in good agreement with theory.

\begin{figure}[htbp]
\centerline{\includegraphics[height=1.65in]{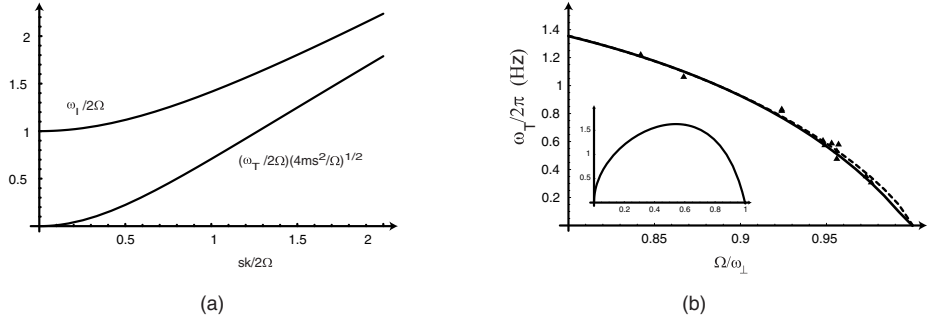}}
\caption {a) Gapped sound and Tkachenko mode frequencies vs. wavevector,
for $\Omega \ll ms^2$.  (b) Frequency of the lowest Tkachenko mode, with the
mean sound velocity, a decreasing shear modulus, $C_2$ (solid curve) and a
constant $C_2$ (dashed curve).  The data (triangles) are from
Ref.~\onlinecite{jilatk}.  The inset shows $\omega_T$ over the entire range of
$\Omega$, at constant $C_2$.}
\label{FIG2}
\end{figure}

    The softness of the Tkachenko modes in the rapidly rotating regime leads
to infrared singular behavior in the vortex transverse
displacement-displacement correlations at finite temperature, and in the order
parameter phase correlations even at zero temperature.\cite{qhmodes,SHM,drew}
The zero point oscillations of the vortices cause quantum melting of the
lattice when $N_v \sim 10-20 N$, where $N_v$ is the total number of vortices
present.\cite{qhmodes,SHM} In a finite system the single particle density
matrix, $\langle\psi(r)\psi^\dagger(r')\rangle$, falls algebraically as $|\vec
r-\vec r\,'|^{-\eta}$, where\cite{qhmodes} $\eta \simeq (ms^2 n/8C_2)^{1/2}
N_v/N$.  Dephasing of the condensate becomes significant only
as $N_v\to N$, and not necessarily before the vortex lattice melts.

    However, in three dimensions, the Tkachenko mode frequency at long
wavelengths becomes linear in the wavevector for any propagation direction out
of the transverse plane.\cite{drew} At zero temperature the vortex
displacement correlations are convergent at large separation, but at finite
temperatures, they grow with separation.  The growth of the vortex
displacements should lead to observable melting of vortex lattices at higher
temperatures and somewhat lower particle number and faster rotation than in
current experiments.  At zero temperature a system of large extent in the
axial direction maintains long range order-parameter correlations for large
separation, but at finite temperatures the correlations decay with
separation.\cite{drew}

\section{BEYOND THE LLL REGIME}

    At sufficiently high rotation, the vortex lattice should melt and become a
vortex liquid.  The regime just beyond melting has yet to be described in
detail.  At still higher rotation speeds, as seen in numerical simulations
with a limited number of particles, the system begins to enter a sequence of
highly correlated incompressible fractional quantum Hall-like
states.\cite{Cooper,Viefers,Read,Jolicoeur} For example, at angular momentum
$L_z = 2N(N-1)$, where $N_v$ (measured in terms of the total circulation,
$N_v=(m/h)\oint \vec v\cdot d\vec\ell$) equals $2N$, the exact ground state
is an $N$-particle fully symmetric Laughlin wave function (in two dimensions),
\begin{equation}
   \Psi(r_1,r_2,\dots,r_N) \sim \prod_{i\ne j}(\zeta_i-\zeta_j)^2 e^{-\Sigma_k
r_k^2/2d_\perp^2},
\end{equation}
where $\zeta_j = x_j+iy_j$.  Since the wave function vanishes whenever two
particles overlap, the interaction energy, $\frac12 g\Sigma_{i \ne
j}\delta(\vec r_i-\vec r_j)$, vanishes in this state.  Theoretically
elucidating the states in general when the angular momentum per particle is of
order the total particle number, as well as studying this regime
experimentally, remain important challenges.

\begin{figure}[htbp]
\centerline{\includegraphics[height=1.5in]{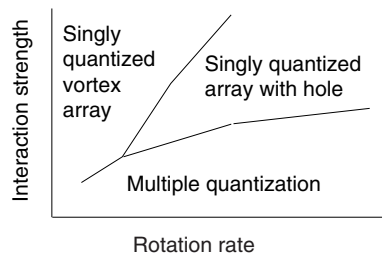}}
\caption
{Schematic phase diagram of the ground state of a rapidly rotating Bose
condensed gas at zero temperature in the $\Omega$--$N a/Z$ plane, where $Z$ is
the height in the $z$ direction, showing the regions of multiply quantized
vortices, of singly quantized vortices forming an array which at large number
becomes a triangular lattice, and of an array of singly quantized vortices
with a hole in the center.}
\label{FIG3}
\end{figure}

\section{ANHARMONIC TRAPS}

    The physics of a condensate confined in an anharmonic trap, e.g.,
(\ref{anharmonic}), is quite different from that in a harmonic trap, since it
becomes possible to rotate the system arbitrarily fast.  As the system rotates
sufficiently rapidly, the centrifugal force pushes the particles towards the
edge of the trap, and a hole open up in the center.  Singly quantized vortex
arrays with a hole have been seen in numerical simulations\cite{tku} and
discussed theoretically in Refs.~\onlinecite{FG,KB,JK,JKL,KBJ}.  In addition,
at very high rotation, systems tend to form a single multiply quantized vortex
at the center, with order parameter $\psi\sim e^{i\nu \phi}$, where the
integer quantization index $\nu$ is $>>$ 1. Such giant vortices have been seen
in numerical simulations\cite{tku,Lundh}, and are discussed theoretically in
Refs.~\onlinecite{FG,KB,JK,JKL}.  The schematic phase diagram, as a function
of interparticle interaction strength vs. rotation rate is shown in Fig. 3.
Full details can be found in Refs.~\onlinecite{KB,JK,JKL,KBJ}.  Initial
studies of rapidly rotating condensates in harmonic lattices at the ENS are
reported in Ref.~\onlinecite{dalibard}.

\section*{ACKNOWLEDGMENTS}

    The research reported here grew out of collaborations with Chris Pethick,
Uwe Fischer, Drew Gifford and Gentaro Watanabe, and was supported in part by
NSF Grants PHY00-98353 and PHY03-55014.  I am very grateful to Eric Cornell,
Ian Coddington, Peter Engels, and Volker Schweikhard of JILA for many
illuminating discussions of their experiments on rapidly rotating Bose
condensates.

\end{document}